\documentclass[epj,twocolumn]{webofc}
\usepackage[varg]{txfonts}   
\usepackage{aas_macros}
\usepackage[pdftex]{hyperref}

\woctitle{The Innermost Regions of Relativistic Jets and Their Magnetic Fields}
\begin{document}
\title{An Exceptional Radio Flare in Markarian 421}
%
%

\author{Joseph~L. Richards\inst{1}\fnsep\thanks{\email{jlr@purdue.edu}} \and
  Talvikki Hovatta\inst{2} \and
  Matthew~L. Lister\inst{1} \and
  Anthony~C.~S. Readhead\inst{2} \and
  Walter Max-Moerbeck\inst{3} \and
  Tuomas Savolainen\inst{4} \and
  Emmanouil Angelakis\inst{4} \and
  Lars Fuhrmann\inst{4} \and
  Margo~F. Aller\inst{5} \and
  Hugh~D. Aller\inst{5} \and
  Ioannis Myserlis\inst{4} \and
  Vassilis Karamanavis\inst{4}
}

\institute{
  Purdue University \and
  California Institute of Technology \and
  NRAO \and
  MPIfR \and
  University of Michigan
}

\hypersetup{
  pdftitle={An Exceptional Radio Flare in Markarian 421},
  pdfauthor={Joseph L. Richards},
  hidelinks
}

\abstract{ In September 2012, the high-synchrotron-peaked (HSP) blazar
  Markarian~421 underwent a rapid wideband radio flare, reaching
  nearly twice the brightest level observed in the centimeter band in
  over three decades of monitoring. In response to this event we
  carried out a five epoch centimeter- to millimeter-band
  multifrequency Very Long Baseline Array (VLBA) campaign to
  investigate the aftermath of this emission event. Rapid radio
  variations are unprecedented in this object and are surprising in an
  HSP BL~Lac object. In this flare, the 15~GHz flux density increased
  with an exponential doubling time of about 9~days, then faded to its
  prior level at a similar rate. This is comparable with the fastest
  large-amplitude centimeter-band radio variability observed in any
  blazar. Similar flux density increases were detected up to
  millimeter bands. This radio flare followed about two months after a
  similarly unprecedented GeV gamma-ray flare (reaching a daily
  ${E>100~\mathrm{MeV}}$ flux of
  ${(1.2\pm0.7)\times10^{-6}~\mathrm{ph\,cm^{-2}\,s^{-1}}}$) reported
  by the Fermi Large Area Telescope (LAT) collaboration, with a
  simultaneous tentative TeV detection by ARGO-YBJ. A
  cross-correlation analysis of long-term 15~GHz and LAT gamma-ray
  light curves finds a statistically significant correlation with the
  radio lagging $\sim\!\!40$~days behind, suggesting that the
  gamma-ray emission originates upstream of the radio emission.
  Preliminary results from our VLBA observations show brightening in
  the unresolved core region and no evidence for apparent superluminal
  motions or substantial flux variations downstream.}
\maketitle
\section{Introduction} \label{intro} Markarian~421 (Mrk~421,
${z=0.031}$) is one of the nearest, best known, and most intensively
studied blazars (including four talks and two posters at this meeting,
\cite{lico_these_proceedings,niinuma_these_proceedings,mastichiadis_these_proceedings,balokovic_these_proceedings,racero_these_proceedings}). It
is classified as a BL~Lac object and was the first extragalactic
gamma-ray source detected at TeV
energies~\cite{punch_detection_1992}. At these energies it is one of
the brightest and fastest-varying extragalactic objects, with a
history of strong variations on timescales of hours or
less~\cite{gaidos_extremely_1996}. It is also a bright X-ray source,
and is frequently detected in GeV gamma-rays.

\begin{figure*}
  \centering
  \includegraphics[height=2.3in]{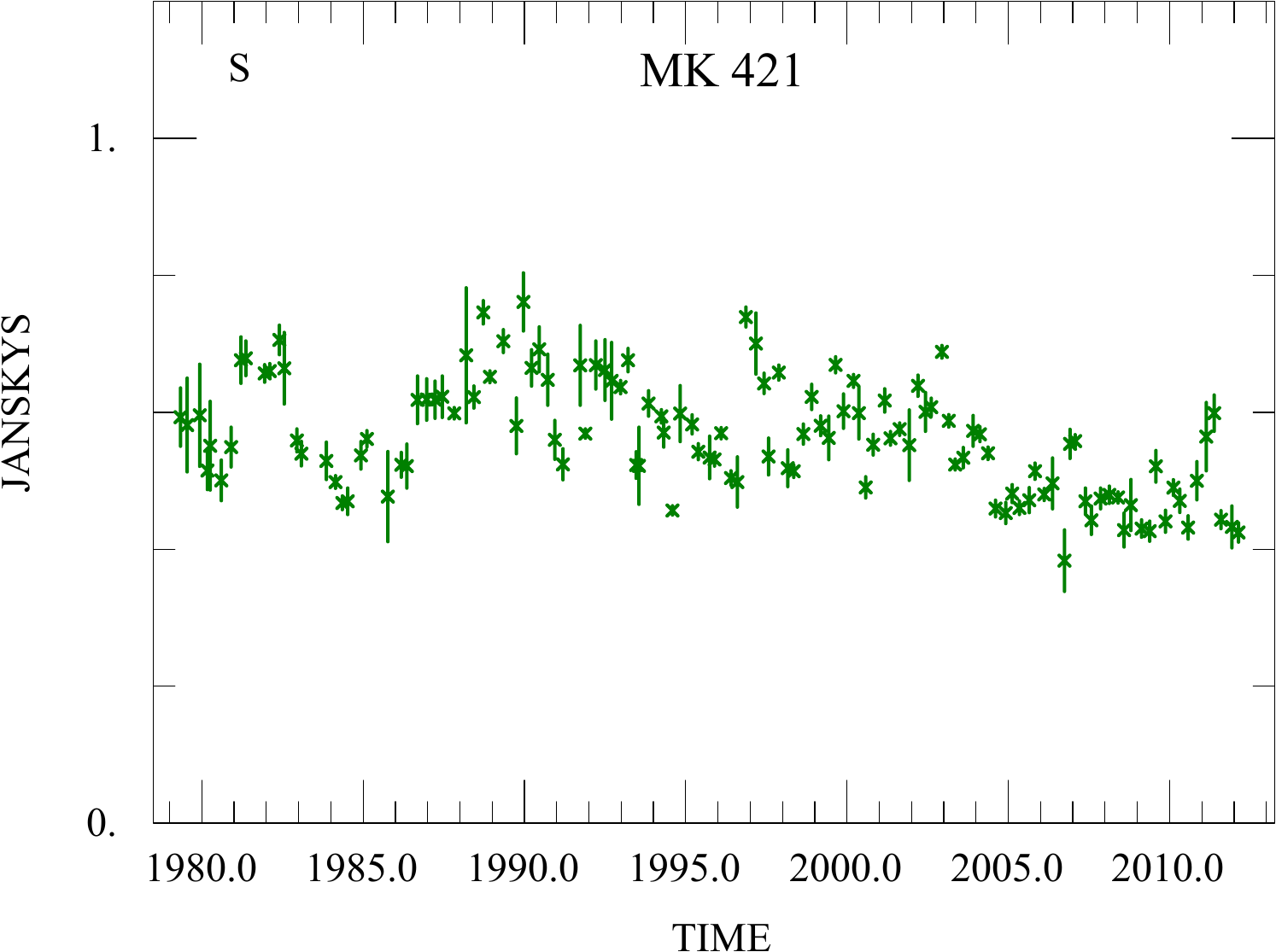} \hfill
  \includegraphics[height=2.4in]{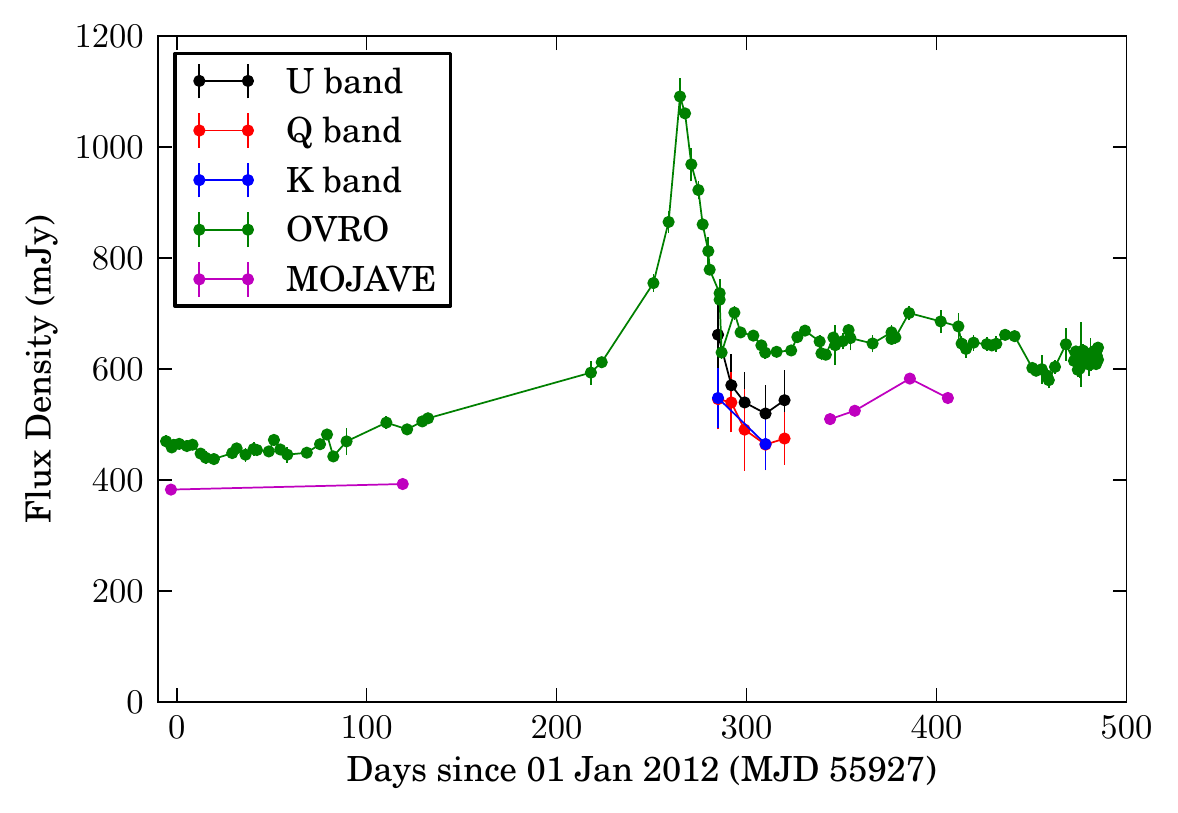}
  \caption{Long-term radio light curves for Mrk~421, including UMRAO
    14.5~GHz data (left panel) and OVRO 15~GHz, MOJAVE 15~GHz, and our
    multifrequency VLBA data (right panel). Vertical scales are equal
    in both panels.}
  \label{fig:longterm_lc}
\end{figure*}

Based on its spectral energy distribution (SED), which shows a
synchrotron peak in the X-ray band, Mrk~421 is classified as a
high-synchrotron-peaked (HSP) object~\cite{abdo_mrk421sed_2011}. Its
radio behavior is rather tame, typical of HSP
blazars~\cite[e.g.,][]{piner_jets_2010,ackermann_2lac}. As illustrated
in the left panel of figure~\ref{fig:longterm_lc}, no major
variability was detected in more than 30 years of regular monitoring
at 14.5~GHz by the University of Michigan Radio Astronomical
Observatory (UMRAO). Extensive studies using very long baseline
interferometry (VLBI) have not found evidence for apparent
superluminal motion in the components in the parsec-scale regions
around the core~\cite[e.g.,][]{piner_polarization_2005,MOJAVE_VI_kinematics,lico_vlba_2012}.

In September 2012, an unprecedented radio flare was detected by the
Owens Valley Radio Observatory (OVRO) 40~m blazar monitoring
program~\cite{richards_et_al_2011, ovro_mrk421_atel_2012}. The OVRO
15~GHz flux densities are plotted in green in the right panel of
figure~\ref{fig:longterm_lc}. In response to this unexpected radio
event, we executed a rapid-cadence five-epoch polarimetric monitoring
campaign with the Very Long Baseline Array (VLBA) to investigate
whether any parsec-scale structural or kinematic changes occurred.  We
present here preliminary results from our U-band total intensity VLBA
observations and from our single-dish monitoring.

\section{Details of the Flare}
The first signs of the radio flare that peaked in September 2012
appeared in mid-August 2012. In late August and September, the OVRO
15~GHz flux density measurements of Mrk~421 increased exponentially
from about 550~mJy to a peak of ${(1.09\pm0.03)}$~Jy on 21~September
2012 with a doubling time of about 9~days. As can be seen in
figure~\ref{fig:longterm_lc}, the flux density then returned to a
lower but still elevated level along a similar exponential
profile. This is extremely rapid radio variation for a blazar. A study
of about 500~flares in about 80~blazars at 22~and 37~GHz, where
variability is typically faster and stronger than at 15~GHz, found a
median doubling time of about 60~days and none faster than
9~days~\cite{hovatta_doppler_2009}.

\begin{figure}
  \centering
  \includegraphics[width=\columnwidth]{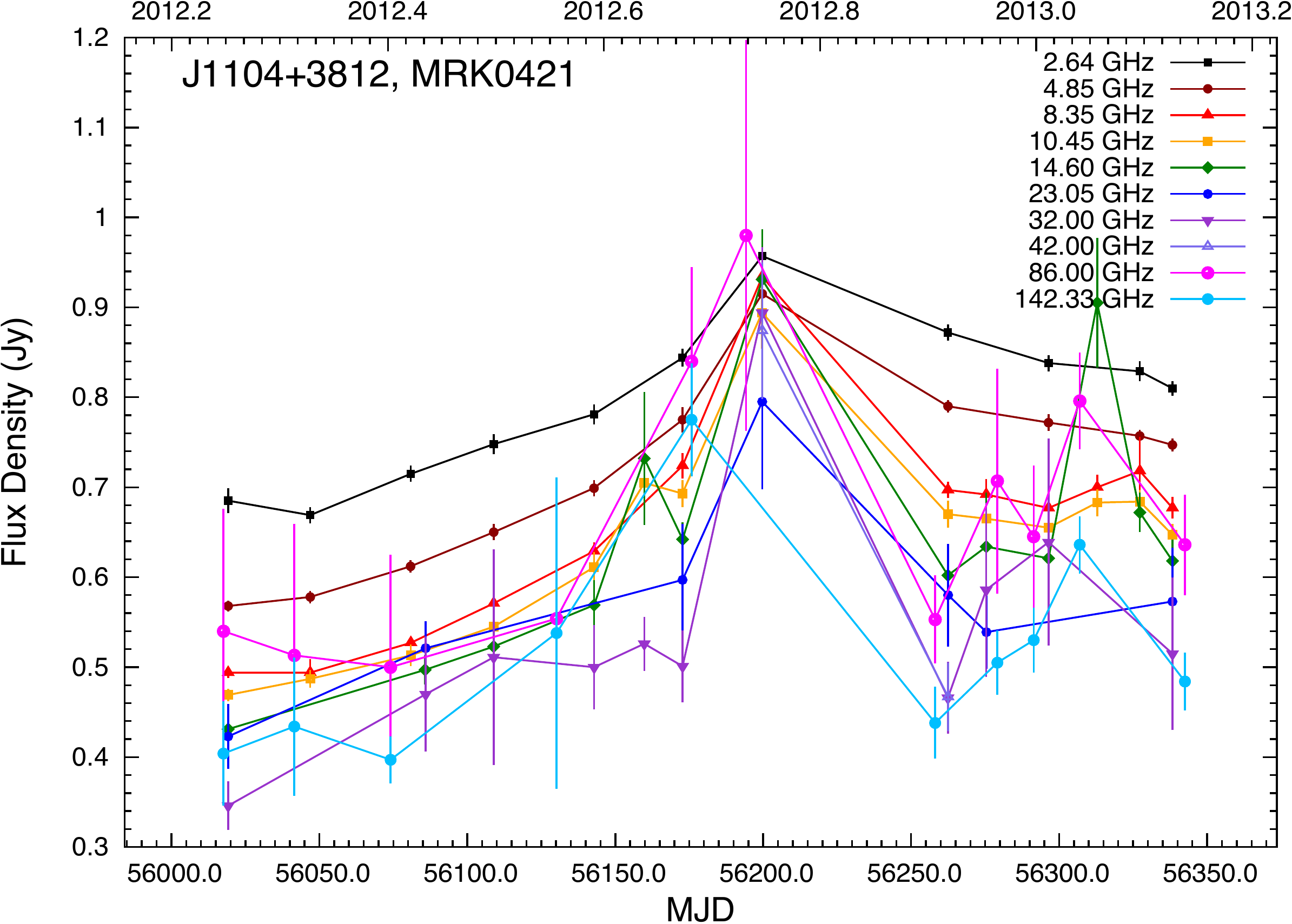}
  \caption{F-GAMMA radio light curves from 2.6~to 142~GHz.}
  \label{fig:radio_spectra}
\end{figure}

The flare at 15~GHz was accompanied by nearly simultaneous flaring at
all observed frequencies from 2.6~to 142~GHz. These multiband radio
observations, shown in figure~\ref{fig:radio_spectra}, were obtained
through the F-GAMMA
program\footnote{\url{http://www3.mpifr-bonn.mpg.de/div/vlbi/fgamma/fgamma.html}}
as part of an ongoing gamma-ray blazar monitoring campaign. The
F-GAMMA light curves appear to show a doubly-peaked flare profile at
several frequencies including 14.6~GHz, with an earlier peak near
16~August 2012, but the uncertainty on these points is high. OVRO data
about a week earlier do not show signs of this early peak.

This multiband radio flare followed shortly after a major GeV
gamma-ray outburst reported by the \emph{Fermi} Large Area Telescope
(LAT) Collaboration~\cite{fermi_mrk421_atel_2012}. A tentative
detection at TeV energies coincident with the GeV flare was also
reported~\cite{argoybj_mrk421_atel_2012}, but other multiwavelength
follow-up during and after this flare was precluded because Mrk~421
was too near the Sun. Later multiwavelength observations of the source
in early 2013 found major variability and flaring in many
bands~\cite[e.g.,][and related ATels]{nustar_mrk421_atel_2013}.

We can estimate the Doppler factor of the jet from the radio
variability, following~\cite{hovatta_doppler_2009}. Assuming the light
travel time across the emission region sets the variability timescale,
we can estimate the brightness temperature of the emission region
during the flare to be
\begin{equation}
  T_{\mathrm{b,var}}=\left(1.548\times10^{-32}~\mathrm{K}\right)\times\frac{\Delta Sd_\mathrm{L}^{2}}{\nu^2\tau^2\left(1+z\right)},
\end{equation}
where ${\Delta S\approx0.5~\mathrm{Jy}}$ is the amplitude of the flare
in janskies,
${d_{\mathrm{L}}=133~\mathrm{Mpc}=4.1\times10^{24}~\mathrm{m}}$ is the
luminosity distance to Mrk~421 in meters, ${\nu=15~\mathrm{GHz}}$ is
the observation frequency in GHz, ${\tau\approx13~\mathrm{days}}$ is
the exponential time constant for the flare in days, and ${z=0.031}$
is the redshift. With these parameters, we find
${T_{\mathrm{b,var}}\approx3.3\times10^{12}~\mathrm{K}}$. Assuming the
rest-frame brightness temperature of the emission region is near the
equipartition limit,
${T_{\mathrm{b,eq}}\approx5\times10^{10}~\mathrm{K}}$~\cite{readhead_equipartition_1994},
we estimate the Doppler factor to be
\begin{equation}
  D_{\mathrm{var}}=\left(\frac{T_{\mathrm{b,var}}}{T_{\mathrm{b,eq}}}\right)^{1/3}\approx4.1.
\end{equation}
This modest value is in accord with previous estimates of the Doppler
and Lorentz factors based on VLBI
data~\cite[e.g.,][]{piner_polarization_2005, lico_vlba_2012}.
Estimates of the Doppler factor from high energy observations find
much higher values,
$D\sim20$~\cite[e.g.,][]{maraschi_simultaneous_1999}. This apparent
conflict thus remains, perhaps indicating a velocity structure in the
jet~\cite[see][for related discussions in these proceedings]{lico_these_proceedings,piner_these_proceedings,lister_these_proceedings}.

\section{VLBA Follow-Up Observations}
In response to the September 2012 extreme radio flare, we undertook a
multiwavelength polarimetric VLBA monitoring campaign to characterize
the structure and kinematics of the parsec-scale jet. The total flux
density at 15~GHz (U~band), 24~GHz (K~band), and 43~GHz (Q~band) are
plotted in figure~\ref{fig:longterm_lc}. The first epoch was observed
on 12~October, 2012, about 21 days after the peak of the 15~GHz
flare. At this time, the flux density had dropped substantially from
its peak.

\begin{figure}
  \centering
  \includegraphics[height=2.8in]{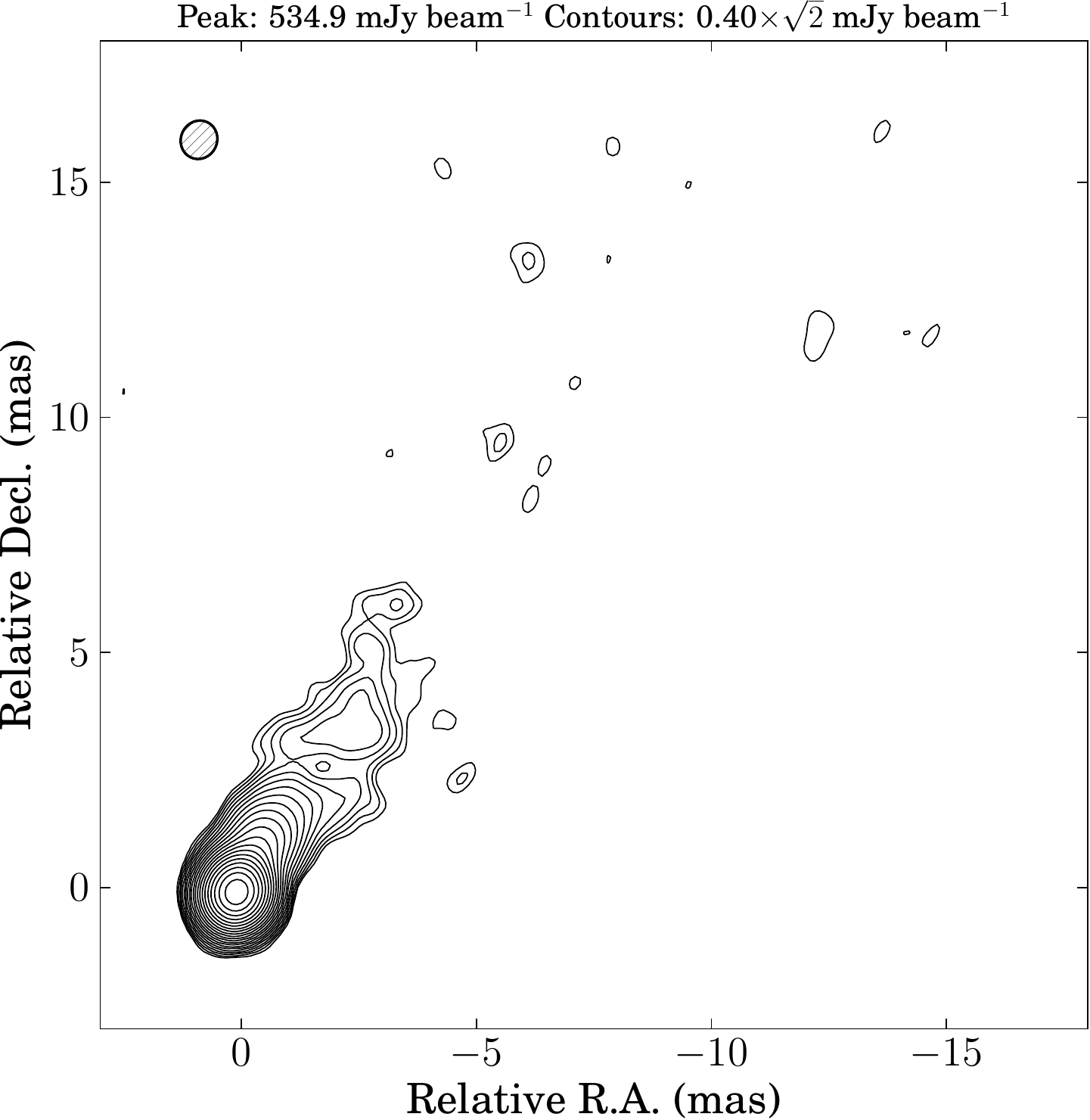}
  \caption{Total intensity contour map from the 12~October 2012 U-band
    (15~GHz) VLBA observation.}
  \label{fig:a_epoch_contours}
\end{figure}

\begin{figure}
  \centering
  \includegraphics[height=2.8in]{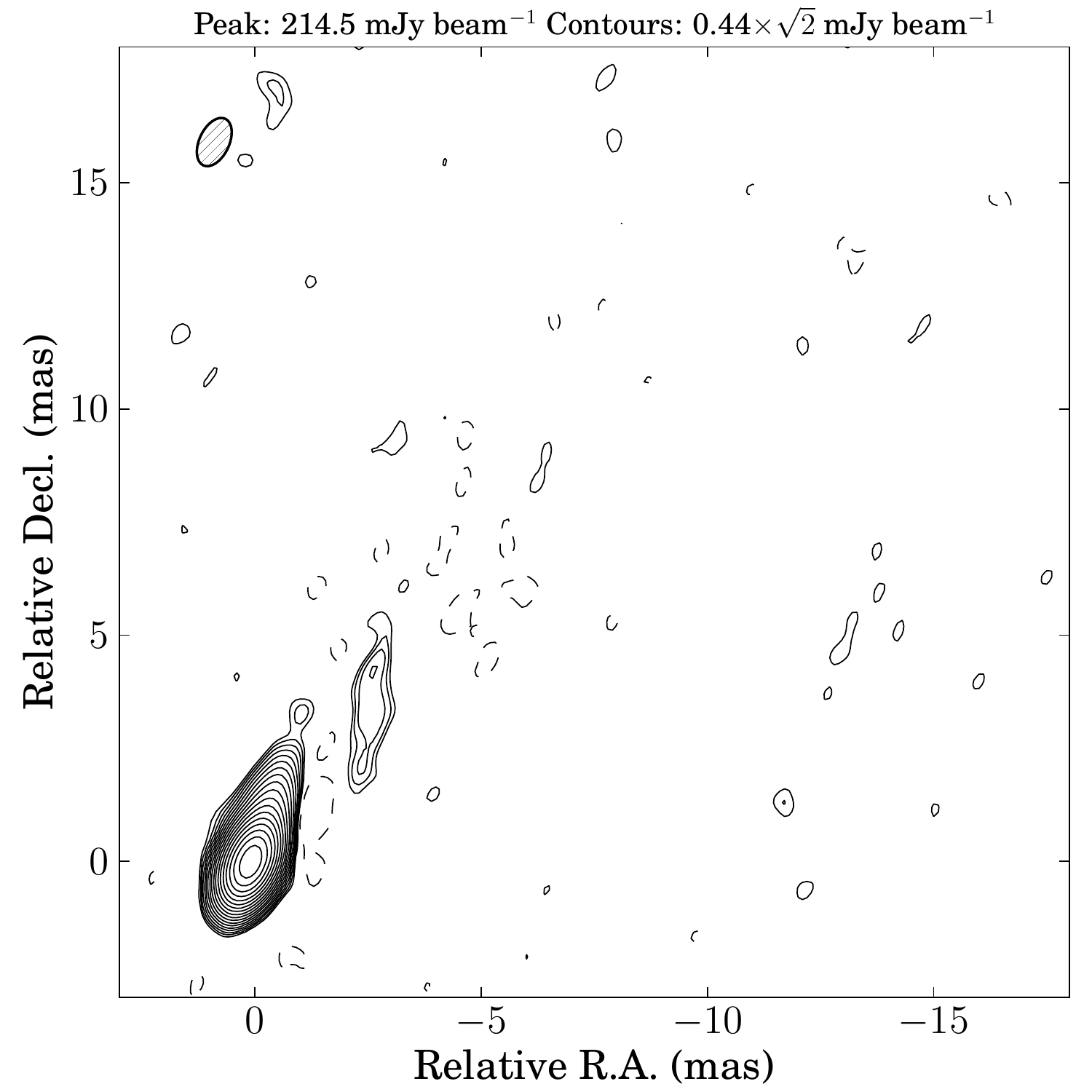}
  \caption{Difference in total intensity between the 12~October 2012
    U-band map and the preceding MOJAVE epoch, observed 29~April
    2012. Before subtracting, both cleaned maps were restored with the
    elliptical Gaussian beam whose position angle and FWHM size is
    shown in the upper left. }
  \label{fig:a_minus_aj}
\end{figure}

We reduced our VLBA data using standard methods with AIPS and
Difmap~\cite{difmap}.  A total intensity contour map for the U-band
data from our first VLBA observation is shown in
figure~\ref{fig:a_epoch_contours}. This map differs little
morphologically from the
MOJAVE\footnote{\url{http://www.physics.purdue.edu/mojave}} U-band map
obtained about five months earlier on 29~April
2012. Figure~\ref{fig:a_minus_aj} shows the difference between these
maps, which is largely confined to the core region. Our K- and Q-band
observations show a very similar morphology. Preliminary analysis of
our subsequent U-band total intensity observations shows a fading of
the core region, but no major structural changes.

\begin{figure}
  \centering
  \includegraphics[height=2.8in]{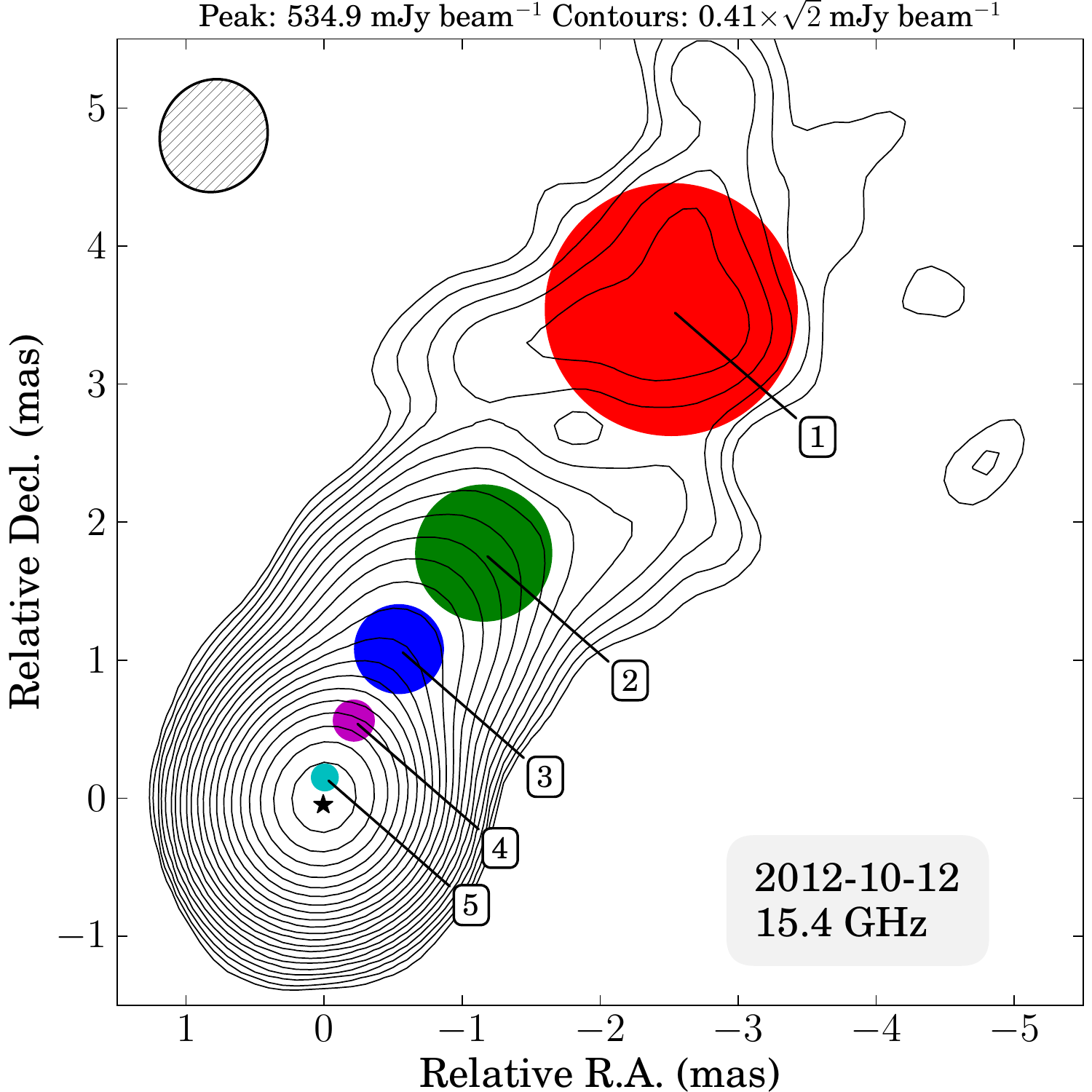}
  \caption{Fitted components overlaid on the 12~October 2012 U-band
    total intensity contour map. Components are circular Gaussians
    with FWHM illustrated by the size of the plotted circles. The core
    is modeled as a delta-function component, plotted as a star.}
  \label{fig:model_components}
\end{figure}

\begin{figure}
  \centering
  \includegraphics[width=\columnwidth]{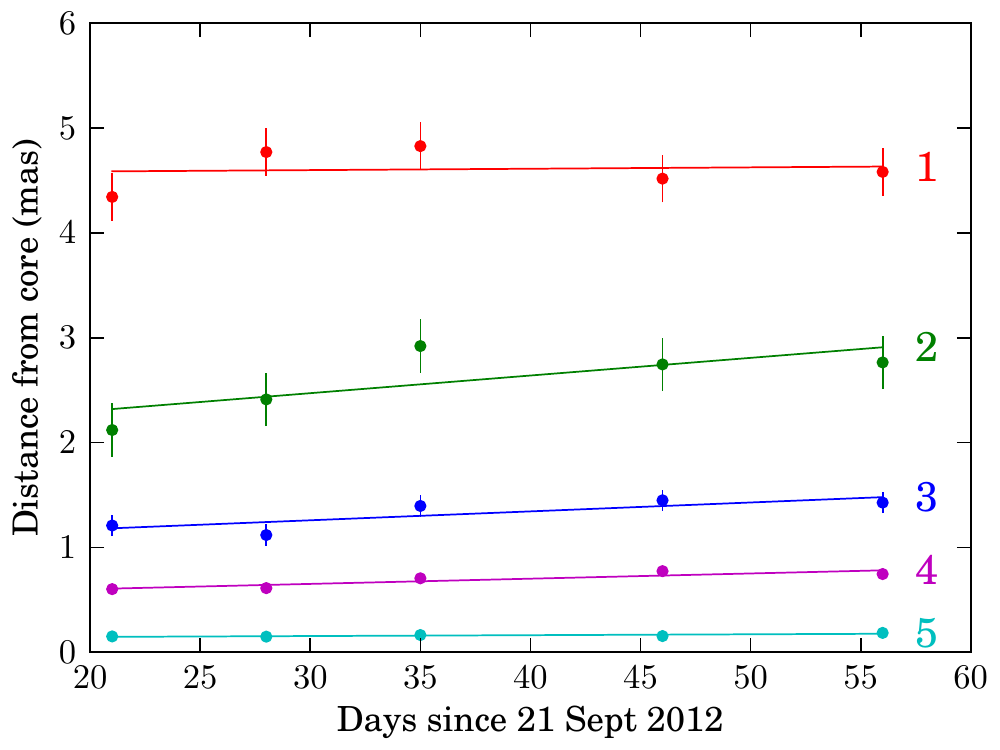}
  \caption{Radial distance of each component in the U-band model fit
    at each epoch.}
  \label{fig:radius_vs_epoch}
\end{figure}

To examine the kinematics of the jet, we used Difmap to fit a model
consisting of an elliptical Gaussian or delta-function core (depending
on epoch) and circular Gaussian downstream components to our 15~GHz
$(u,v)$ data. The model for the first epoch is shown in
figure~\ref{fig:model_components}. We obtained reasonable reduced
$\chi^2$ in all epochs using a six-component model. The radial
distance from each downstream component to the core component in all
five epochs is plotted in figure~\ref{fig:radius_vs_epoch}. By fitting
a linear trend for each component, we determined their apparent
speeds, tabulated in table~\ref{tab:speeds}. All are consistent (to
$2\sigma$) with stationarity or subluminal speed.

\begin{table}
  \centering
  \caption{Apparent component speeds}
  \label{tab:speeds}
  \begin{tabular}{c c} \hline
    Component & Apparent Speed \\\hline
    1 & $(0.3\pm1.8)c$ \\
    2 & $(3.8\pm2.1)c$ \\
    3 & $(1.9\pm0.9)c$ \\
    4 & $(1.2\pm0.4)c$ \\
    5 & $(0.2\pm0.1)c$ \\\hline
  \end{tabular}
\end{table}

\section{Radio/Gamma-ray Cross Correlation}

One must exercise caution when concluding that a physical connection
exists between individual flares based solely on coincidence in
time. Because of the extremely rare nature of the radio flare and the
infrequency of gamma-ray flares, this particular case is more
convincing than many. Still, a quantitative estimate of the
significance of the occurrence of two flares is desirable. To
estimate significances, we have performed a set of Monte Carlo
simulations.

\begin{figure}
  \centering
  \includegraphics[width=\columnwidth]{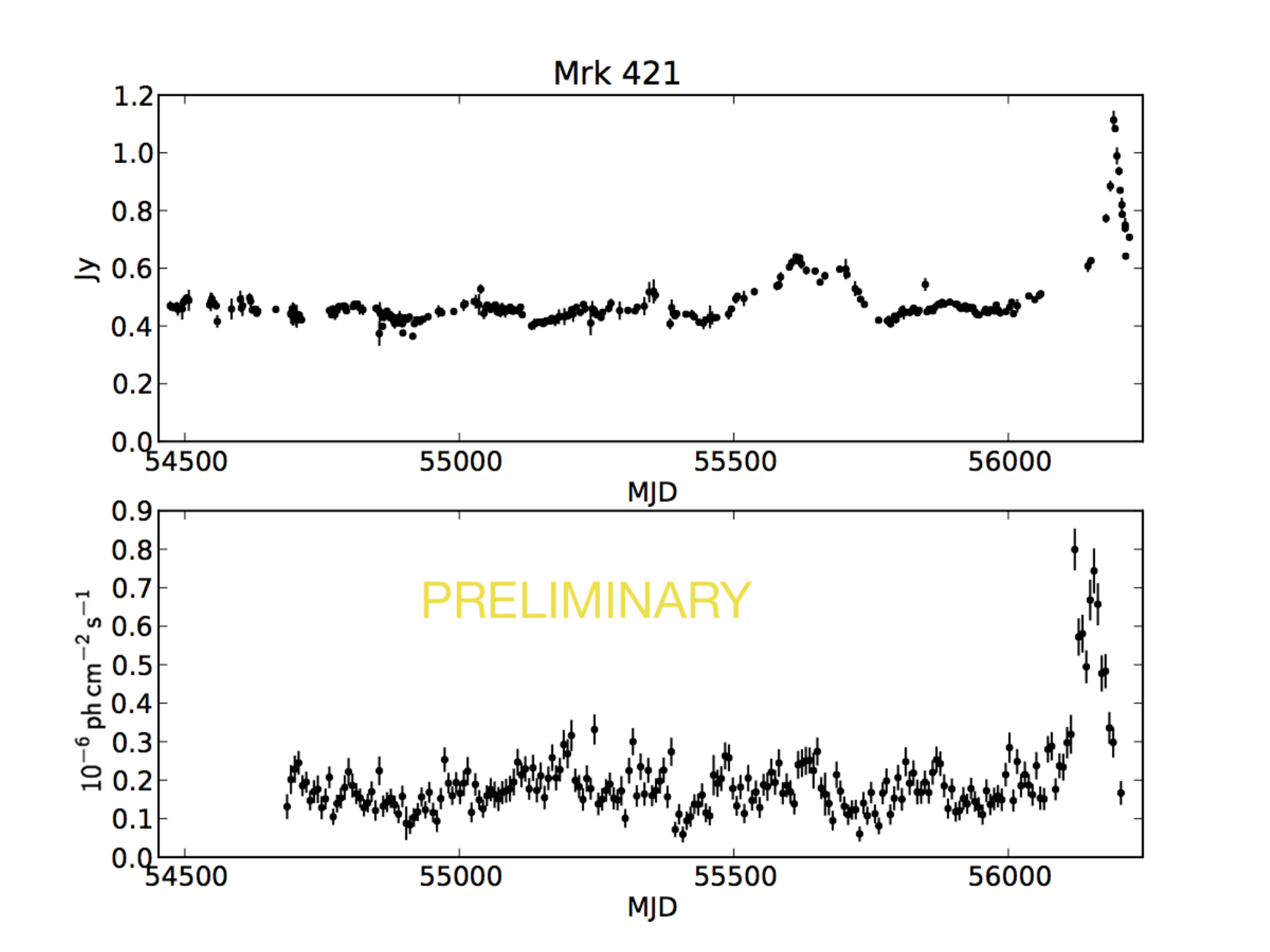}
  \caption{OVRO 40~m 15~GHz radio light curve (top) and
    \emph{Fermi}-LAT 100~MeV--200~GeV weekly binned gamma-ray light
    curve for Mrk~421.}
  \label{fig:gamma_radio_lc}
\end{figure}

\begin{figure}
  \centering
  \includegraphics[width=\columnwidth]{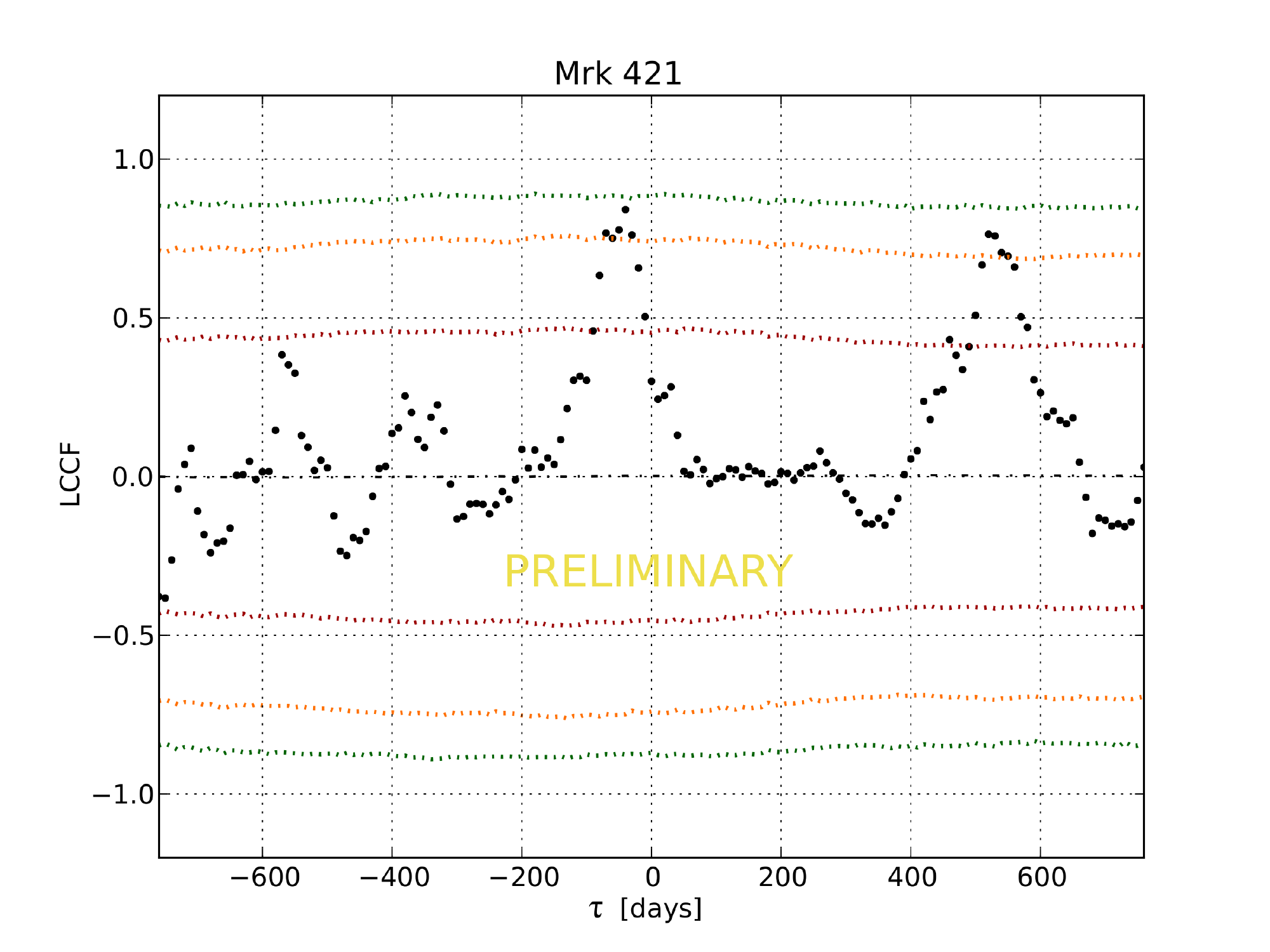}
  \caption{Discrete cross-correlation function between the radio and
    gamma-ray light curves shown in
    figure~\ref{fig:gamma_radio_lc}. Dashed lines indicate the
    positive and negative $1\sigma$, $2\sigma$, and $3\sigma$
    significance contours obtained through Monte Carlo simulation.}
  \label{fig:cross_correlation}
\end{figure}

In figure~\ref{fig:gamma_radio_lc}, the preliminary weekly-binned LAT
light curve and the OVRO 15~GHz light curves are shown. The discrete
cross-correlation function (DCF) of these two light curves, computed
following~\cite{edelson_and_krolik_1988} to account for uneven
sampling, is shown in figure~\ref{fig:cross_correlation}. The positive
and negative contours show the $1\sigma$, $2\sigma$, and $3\sigma$
significance levels. These are determined by simulating 20,000 mock
radio and gamma-ray light curves using the same uneven sampling as the
original data sets. The light curves are assumed to result from red
noise processes with noise power spectral densities proportional to
$f^{-\beta}$, using the true light curves to estimate $\beta$
separately for each band.  Further details of this method and a
previous discussion of this Mrk~421 event are found
in~\cite{max-moerbeck_ovro_2013}.

The DCF peak we find is nearly at the $3\sigma$ significance
level. Very few blazars in the OVRO monitoring program reach this
level of significance according to our method. If we exclude the
flaring period, the significance of the highest peak drops below
$1\sigma$. Thus, our method agrees with our intuitive sense that these
light curves show a true correlation. However, there is ambiguity in
this result: there are two peaks in the DCF, both of similar amplitude
and significance. The peak at $-40$~days results from the overlap of
the September radio flare with the gamma-ray flare. The second peak,
near a lag of 500~days, results from the overlap of the smaller radio
high state apparent near MJD~55650 in the radio light curve in
figure~\ref{fig:gamma_radio_lc}. Either pairing of events is unlikely
to occur by chance, but with only a single major feature in the
gamma-ray light curve we cannot unambiguously determine the lag from
the light curves alone. Because of their similar appearances in the
plot, it is tempting to conclude that the larger radio flare
corresponds to the gamma-ray flare, but this implicitly assumes a
model for the physical process connecting the emission in the two
bands.

It is important to note that for this test we explicitly selected a
time interval including the flaring period in both bands. As such, our
statistical result is \emph{a posteriori}, and does not represent
independent statistical confirmation of this correlation. To claim
such a confirmation, we would require the comparison of long light
curves with a time interval selected independently of source
activity. For now, the chief evidence supporting this is the temporal
coincidence of rare events.

\section{Conclusions}
Mrk~421 displayed a radio emission event unique in its long radio
monitoring history. The small delay between this extremely rare event
and a preceding GeV and possibly TeV gamma-ray flare, both infrequent
occurrences in this source, provides intuitively compelling evidence
for a physical connection between them. Our quantitative method for
finding significant correlations between light curves supports this
conclusion, although it does not provide independent evidence. If the
gamma-ray flare and the September~2012 radio flare are, in fact,
related, there is a delay of about 40~days from the gamma-ray flare to
the radio flare.  Preliminary analysis of the 15~GHz radio light curve
and our VLBA follow-up observations does not find evidence for the
ejection of a superluminal component. Our data are consistent with
previous radio determinations of the jet Doppler factor,
${D\approx3}$. Thus, there remains a conflict between this and the
much higher Doppler factors required by the high-energy emission. It
seems that if this is indicative of velocity structure, even in this
unusual radio flare, we did not detect radio emission from a putative
faster region of the jet.

\begin{acknowledgement}
J.~R. and T.~H. acknowledge the support of the American Astronomical
Society and the National Science Foundation (NSF) in the form of
International Travel Grants. T.~H. was supported by the Jenny \& Antti
Wihuri foundation. This work and the MOJAVE program were supported by
NASA \emph{Fermi} Guest Investigator grant 11-Fermi11-0019. The OVRO
40~m monitoring program is supported in part by NSF grants AST-0808050
and AST-1109911 and NASA grants NNX08AW31G and NNX11AO43G. UMRAO was
supported in part by NSF grant AST-0607523 and NASA \emph{Fermi} Guest
Investigator grants NNX09AU16G, NNX10AP16G, and NNX11AO13G.  The
National Radio Astronomy Observatory is a facility of the National
Science Foundation operated under cooperative agreement by Associated
Universities, Inc.
\end{acknowledgement}


%
\bibliography{richards}

\begin{thebibliography}{26}

\bibitem{lico_these_proceedings}
R.L. Lico et~al., these proceedings  (2013)

\bibitem{niinuma_these_proceedings}
K.~Niinuma et~al., these proceedings  (2013)

\bibitem{mastichiadis_these_proceedings}
A.~Mastichiadis, M.~Petropoulou, S.~Dimitrakoudis, these proceedings  (2013)

\bibitem{balokovic_these_proceedings}
M.~Balokovi\'{c} et~al., these proceedings  (2013)

\bibitem{racero_these_proceedings}
E.~Racero, I.~de~la Calle, these proceedings  (2013)

\bibitem{punch_detection_1992}
M.~{Punch} et~al., \nat{} \textbf{358}, 477 (1992)

\bibitem{gaidos_extremely_1996}
J.A. {Gaidos} et~al., \nat{} \textbf{383}, 319 (1996)

\bibitem{abdo_mrk421sed_2011}
A.A. {Abdo} et~al., \apj{} \textbf{736}, 131 (2011)

\bibitem{piner_jets_2010}
B.G. {Piner}, N.~{Pant}, P.G. {Edwards}, \apj{} \textbf{723}, 1150 (2010)

\bibitem{ackermann_2lac}
M.~{Ackermann} et~al., \apj{} \textbf{743}, 171 (2011)

\bibitem{piner_polarization_2005}
B.G. Piner, P.G. Edwards, \apj{} \textbf{622}, 168 (2005)

\bibitem{MOJAVE_VI_kinematics}
M.L. {Lister} et~al., \aj{} \textbf{138}, 1874 (2009)

\bibitem{lico_vlba_2012}
R.~{Lico} et~al., \aap{} \textbf{545}, A117 (2012)

\bibitem{richards_et_al_2011}
J.L. {Richards} et~al., \apjs{} \textbf{194}, 29 (2011)

\bibitem{ovro_mrk421_atel_2012}
T.~{Hovatta} et~al., The Astronomer's Telegram \#4451  (2012)

\bibitem{hovatta_doppler_2009}
T.~{Hovatta} et~al., \aap{} \textbf{494}, 527 (2009)

\bibitem{fermi_mrk421_atel_2012}
F.~{D'Ammando} et~al., The Astronomer's Telegram \#4261  (2012)

\bibitem{argoybj_mrk421_atel_2012}
B.~{Bartoli} et~al., The Astronomer's Telegram \#4272  (2012)

\bibitem{nustar_mrk421_atel_2013}
M.~{Balokovi\'{c}} et~al., The Astronomer's Telegram \#4974  (2013)

\bibitem{readhead_equipartition_1994}
A.C.S. {Readhead}, \apj{} \textbf{426}, 51 (1994)

\bibitem{maraschi_simultaneous_1999}
L.~{Maraschi} et~al., \apjl{} \textbf{526}, L81 (1999)

\bibitem{piner_these_proceedings}
G.~Piner, P.G. Edwards, these proceedings  (2013)

\bibitem{lister_these_proceedings}
M.L. Lister et~al., these proceedings  (2013)

\bibitem{difmap}
M.C. {Shepherd}, T.J. {Pearson}, G.B. {Taylor}, \emph{{DIFMAP: an interactive
  program for synthesis imaging.}}, in \emph{Bulletin of the American
  Astronomical Society} (1994), Vol.~26, pp. 987--989

\bibitem{edelson_and_krolik_1988}
R.A. Edelson, J.H. Krolik, \apj{} \textbf{333}, 646 (1988)

\bibitem{max-moerbeck_ovro_2013}
W.~{Max-Moerbeck} et~al., \emph{OVRO 40 m Blazar Monitoring Program: Location
  of the gamma-ray emission region in blazars by the study of correlated
  variability at radio and gamma-rays}, in \emph{Fourth International Fermi
  Symposium Proceedings}, edited by T.J. Brandt, N.~Omodei, C.~Wilson-Hodge
  (eConf C121028, 2013), p. 252, [astro-ph:1303.2131]

\end{thebibliography}
%
%
%
%

\end{document}